\documentclass[aps,twocolumn,showpacs,preprintnumbers,floats]{revtex4}\def\@cite#1#2{\textsuperscript{[{#1\if@tempswa , #2\fi}]}}
\usepackage{mathrsfs}

\usepackage{longtable,lscape}
\usepackage{savesym}
\usepackage{txfonts}
\savesymbol{iint}
\savesymbol{iiint}
\savesymbol{iiiint}
\savesymbol{idotsint}
\usepackage{amssymb}
\usepackage{amsmath}
\restoresymbol{TXF}{iint}
\restoresymbol{TXF}{iiint}
\restoresymbol{TXF}{iiiint}
\restoresymbol{TXF}{idotsint}
\usepackage{indentfirst}
\usepackage{graphicx,booktabs}
\usepackage{braket}
\usepackage{multirow}
\usepackage{color}
\usepackage[colorlinks, citecolor=blue,anchorcolor=red,menucolor=red, linkcolor=red,filecolor=red,urlcolor=blue,frenchlinks=red]{hyperref}
\usepackage{epsfig}
\newcommand{\vsig}{\mbox{\boldmath$\sigma$\unboldmath}}

\begin{document}
	 \bibliographystyle{plain}

	\title{The $1D$-wave bottom-strange baryons and possible interpretation of $\Xi_{b}(6327)^{0}$ and $\Xi_{b}(6333)^{0}$ }
\author{Wen-Jia Wang$^{1}$, Yu-Hui Zhou$^{1}$, Li-Ye Xiao$^{1}$~\footnote {E-mail: lyxiao@ustb.edu.cn}, Xian-Hui Zhong$^{2,3}$~\footnote {E-mail: zhongxh@hunnu.edu.cn}}
\affiliation{ 1)Institute of Theoretical Physics, University of Science and Technology Beijing,
Beijing 100083, China}
\affiliation{ 2) Department of
Physics, Hunan Normal University, and Key Laboratory of
Low-Dimensional Quantum Structures and Quantum Control of Ministry
of Education, Changsha 410081, China }
\affiliation{ 3) Synergetic
Innovation Center for Quantum Effects and Applications (SICQEA),
Hunan Normal University, Changsha 410081, China}

\begin{abstract}

Inspired by the LHCb's newest observation of two new excited $\Xi_b^0$ states, we systematically study the strong decays of the low-lying $\lambda$- and $\rho$-modes $1D$-wave $\Xi_{b}$ and  $\Xi^{'}_{b}$ baryons using the chiral quark model within the $j$-$j$ coupling scheme. Based on the measured masses and strong decay properties of $\Xi_{b}(6327)^{0}$ and $\Xi_{b}(6333)^{0}$, we explain the two states as the $\lambda$-mode $1D$ $\Xi_{b}$ states with $ J^{P}=3/2^{+} $ and $ J^{P}=5/2^{+} $, respectively. Moreover, under this assignment, another dominant decay channel of $\Xi_{b}(6327)^{0}$ is $\Xi'_b\pi$ and that of $\Xi_{b}(6333)^{0}$ is $\Xi_b^*\pi$. Hence, the decay modes $\Xi'_b\pi$ and $\Xi_b^*\pi$ may be another ideal channels as well to decode the inner structure of $\Xi_{b}(6327)^{0}$ and $\Xi_{b}(6333)^{0}$, respectively.
For other unseen $1D$ $\Xi_b$ and $\Xi'_b$ states, our results indicate: (i) $\Xi_b|J^P=\frac{3}{2}^+,2\rangle_{\rho}$ and $\Xi_b|J^P=\frac{5}{2}^+,2\rangle_{\rho}$ are most likely to be narrow states with a width of $\Gamma\simeq(12-30)$ MeV, and dominantly decay into $\Sigma_bK$ and $\Sigma^*_bK$, respectively; (ii) The $1D$ $\Xi'_b$ baryons are not broad states, and the widths vary in the range of $\Gamma\simeq(14-46)$ MeV.  These states have a good potential to be observed in their dominant decay processes.

\end{abstract}
\maketitle

\section{Introduction}

Establishing and improving hadron spectroscopy always is a key subject in hadron physics. Completing this subject can help us to
understand the hadron structure, and then improve our understanding of the dynamics of Quantum Chromodynamics (QCD). As an indispensable part of hadrons, single heavy baryons play an important role since heavy quark symmetry is a good approximation and can provide some qualitative properties, especially in the baryons containing the bottom ($b$) quark. However, differing from the charmed baryons, searching for the bottom states is quite difficult for experiment since higher energy and higher luminance of the beams are required to produce them. Fortunately, experimenters have made important progress in searching for the bottom baryons in recent years~\cite{ParticleDataGroup:2020ssz}, which provides us good opportunities to establish an abundant spectrum by decoding the inner structures of these newly observed bottom baryons.

In 2012, two narrow $1P$ $\Lambda_b^0$ baryons, denoted as $\Lambda_b(5912)^0$ and $\Lambda_b(5920)^0$, were firstly observed by the LHCb Collaboration~\cite{LHCb:2012kxf} and confirmed by the CDF Collaboration~\cite{CDF:2013pvu} the following year. Later, two bottom baryons, i.e., $\Xi_b(6227)^-$~\cite{LHCb:2018vuc} and $\Sigma_b(6097)^{\pm}$~\cite{LHCb:2018haf}, were found by the LHCb Collaboration. Recently, two $1D$ $\Lambda_b^0$ candidates, $\Lambda_b(6146)^0$ and $\Lambda_b(6152)^0$, were discovered by the LHCb Collaboration in the $\Lambda_b^0\pi^+\pi^-$ spectrum~\cite{LHCb:2019soc}. This may be the first time that the low-lying $D$-wave singly bottom baryons are observed in experiment. In addition, four extremely narrow excited $\Omega_{b}$
states, $\Omega_{b}(6316)^{-}$, $\Omega_{b}(6330)^{-}$, $\Omega_{b}(6340)^{-}$  and $\Omega_{b}(6350)^{-}$, were announced by the LHCb Collaboration in the $\Xi_b^0K^-$ mass spectrum\cite{LHCb:2020tqd} in 2020. These observed single bottom baryons have stimulated a wide discussion~\cite{Wang:2018fjm,Wang:2017kfr,Wang:2019uaj,Xiao:2020gjo,Chen:2016spr,Oudichhya:2021yln,
Yu:2021zvl,Kakadiya:2021jtv,Yang:2020zrh,Mao:2020jln,Xiao:2020oif,Lu:2019rtg,Chen:2019ywy,Yao:2018jmc,He:2021xrh,
Jia:2019bkr,Chen:2018orb,Mao:2015gya,Cui:2019dzj}.

Very recently, the LHCb Collaboration again reported their new discovery of two new excited $\Xi_b^0$ states in the $\Lambda_b^0K^-\pi^+$ mass spectrum using a data sample of $pp$ collisions~\cite{LHCb:2021ssn}. The measured masses and decay widths of these two states are
\begin{eqnarray}
m( \Xi_{b}(6327)^{0} )=6327.28^{+0.23}_{-0.21}\pm0.08\pm0.24~\text{MeV},\\
	m( \Xi_{b}(6333)^{0} )=6332.69^{+0.17}_{-0.18}\pm0.03\pm0.22~\text{MeV},
\end{eqnarray}
\begin{eqnarray}
\Gamma( \Xi_{b}(6327)^{0} )<2.20(2.56)~\text{MeV},\\
\Gamma( \Xi_{b}(6333)^{0} )<1.55(1.85)~\text{MeV},
\end{eqnarray}
where the natural widths $\Gamma$ correspond to $90\%(95\%)$ confidence level upper limits. It is also found that
the $\Xi_{b}(6327)^{0}$ resonance observed in the $\Lambda_b^0K^-\pi^+$
final states is dominantly contributed by the intermediate channel $\Sigma_b^{+}K^-$, while
the $\Xi_{b}(6333)^{0}$ resonance is significantly contributed by the intermediate channel $\Sigma_b^{*+}K^-$. To decode their inner structure and analysis their dynamic mechanism in theory is necessary. Before the LHCb's measurement~\cite{LHCb:2021ssn}, there exist many theoretical predictions of the mass spectra of the $\Xi_b$ and $\Xi_b'$ baryons with various models in the literature~\cite{Jia:2019bkr,Kakadiya:2021jtv,Yoshida:2015tia,Chen:2018orb,Mao:2015gya,Yu:2021zvl,Cui:2019dzj,Ebert:2011kk,Roberts:2007ni,
Valcarce:2008dr,Ebert:2007nw,Thakkar:2016dna,Narodetskii:2008pn}, etc. We collect the some theoretical predictions of the spectrum for the $1D$ $\Xi_b$ and $\Xi_b'$ baryons in Table~\ref{table-I}. From the table, the two new exited $\Xi_b^0$ states observed by the LHCb Collaboration
~\cite{LHCb:2021ssn} are in the predicted mass region of the $\lambda$-mode $1D$ $\Xi_b$ resonances with spin-parity $J^P=3/2^+$ and $J^P=5/2^+$~\cite{Chen:2019ywy,Ebert:2011kk,Roberts:2007ni,
Valcarce:2008dr,Ebert:2007nw,Thakkar:2016dna}.
%In addition, the possibility as $\lambda$-mode $1D$ $\Xi'_b$ states with spin-parity $J^P=5/2^+$ and $J^P=7/2^+$ cannot be excluded absolutely based simply on the predicted masses~\cite{Ebert:2007nw}.
Except mass spectrum, decay property is one of the important bases for determining hadron's properties. However, there are only a few discussions of the strong decays of the $1D$ bottom baryons~\cite{Chen:2019ywy,Yao:2018jmc,He:2021xrh}. It should be pointed out that the predicted strong decay properties of the $\lambda$-mode $1D$ $\Xi_b$ resonances with spin-parity $J^P=3/2^+$ and $J^P=5/2^+$ in Refs.~\cite{Chen:2019ywy,Yao:2018jmc} are in good agreement with the properties of the two new exited $\Xi_b^0$ states observed by the LHCb Collaboration~\cite{LHCb:2021ssn}. Recently, Bijker et al.~\cite{Bijker:2020tns} assigned the newly observed $\Xi_{b}(6327)^{0}$ and $\Xi_{b}(6333)^{0}$ states as the $\lambda$-mode $1D$ $\Xi_b$ resonances with spin-parity $J^P=3/2^+$ and $J^P=5/2^+$ as well both within the elementary emission model and the quark-pair creation model. The same theoretical results were obtained in Ref.~\cite{Yu:2021zvl} with the method of QCD sum rules.

\begin{table*}
	\caption{The classifications and masses(MeV) of the $1D$-wave $\Xi_{b}$ and $\Xi^{'}_{b}$ states within the $j$-$j$ coupling scheme.
$\Xi_b$ and $\Xi_b'$ stand for the bottom baryons belonging to flavor antitriplet $\mathbf{\bar{3}}_F$ and flavor sextet $\mathbf{6}_F$, respectively.} \centering \label{table-I}
	\begin{tabular}{ccccccccccccccccccccc}
		\hline\hline
State~~&\multicolumn{6}{c}{Quantum number}~& &\multicolumn{6}{c}{Mass}~~&\\  \hline \cline{2-7}  \cline{9-14}
  $\Xi^{(')}_b|J^P,j\rangle_{\lambda(\rho)}$~~&$l_\lambda$ ~&$l_{\rho}$ ~  &$L$  ~ &$s_{\rho}$ ~  &$j$ ~ &$J^{P}$ ~~&~~~~~  &RQM~\cite{Ebert:2011kk} ~&QM~\cite{Roberts:2007ni}~ &CQC~\cite{Valcarce:2008dr}~&CQM~\cite{Ebert:2007nw}~&hCQM~\cite{Thakkar:2016dna}~&QPM~\cite{Chen:2019ywy}\\ \hline
$\Xi_b|J^P=\frac{3}{2}^+,2\rangle_{\lambda}$ ~~ &2 ~&0~ &2~ &0 ~&2~ &$\frac{3}{2}^+$&  ~~&6366~&6311~&6373~&6359~&6386~&6327\\
$\Xi_b|J^P=\frac{5}{2}^+,2\rangle_{\lambda}$ ~~ &2 ~&0~ &2~ &0 ~&2~ &$\frac{5}{2}^+$& ~~&6373~&6300~&~&6365~&6369~&6330\\ \hline
$\Xi_b|J^P=\frac{3}{2}^+,2\rangle_{\rho}$ ~~    &0 ~&2~ &2~ &0 ~&2~ &$\frac{3}{2}^+$& ~~&~&~&~&~&~&\\
$\Xi_b|J^P=\frac{5}{2}^+,2\rangle_{\rho}$ ~~    &0 ~&2~ &2~ &0 ~&2~ &$\frac{5}{2}^+$& ~~&~&~&~&~&~&\\ \hline
$\Xi'_b|J^P=\frac{1}{2}^+,1\rangle_{\lambda}$ ~~&2 ~&0~ &2~ &1 ~&1~ &$\frac{1}{2}^+$& ~~&6447~&~&~&6420~&~&6486\\
$\Xi'_b|J^P=\frac{3}{2}^+,1\rangle_{\lambda}$ ~~&2 ~&0~ &2~ &1 ~&1~ &$\frac{3}{2}^+$& ~~&6459~&~&~&6410~&~&6488\\
$\Xi'_b|J^P=\frac{3}{2}^+,2\rangle_{\lambda}$ ~~&2 ~&0~ &2~ &1 ~&2~ &$\frac{3}{2}^+$& ~~&6431~&~&~&6412~&~&6456\\
$\Xi'_b|J^P=\frac{5}{2}^+,2\rangle_{\lambda}$ ~~&2 ~&0~ &2~ &1 ~&2~ &$\frac{5}{2}^+$& ~~&6432~&6402~&~&6403~&~&6457\\
$\Xi'_b|J^P=\frac{5}{2}^+,3\rangle_{\lambda}$ ~~&2 ~&0~ &2~ &1 ~&3~ &$\frac{5}{2}^+$& ~~&6420~&~&~&6377~&~&6407\\
$\Xi'_b|J^P=\frac{7}{2}^+,3\rangle_{\lambda}$ ~~&2 ~&0~ &2~ &1 ~&3~ &$\frac{7}{2}^+$& ~~&6414~&6405~&~&6390~&~&6408\\ \hline
$\Xi'_b|J^P=\frac{1}{2}^+,1\rangle_{\rho}$ ~~   &0 ~&2~ &2~ &1 ~&1~ &$\frac{1}{2}^+$& ~~&~&~&~&~&~&\\
$\Xi'_b|J^P=\frac{3}{2}^+,1\rangle_{\rho}$ ~~   &0 ~&2~ &2~ &1 ~&1~ &$\frac{3}{2}^+$& ~~&~&~&~&~&~&\\
$\Xi'_b|J^P=\frac{3}{2}^+,2\rangle_{\rho}$ ~~   &0 ~&2~ &2~ &1 ~&2~ &$\frac{3}{2}^+$& ~~&~&~&~&~&~&\\
$\Xi'_b|J^P=\frac{5}{2}^+,2\rangle_{\rho}$ ~~   &0 ~&2~ &2~ &1 ~&2~ &$\frac{5}{2}^+$& ~~&~&~&~&~&~&\\
$\Xi'_b|J^P=\frac{5}{2}^+,3\rangle_{\rho}$ ~~   &0 ~&2~ &2~ &1 ~&3~ &$\frac{5}{2}^+$& ~~&~&~&~&~&~&\\
$\Xi'_b|J^P=\frac{7}{2}^+,3\rangle_{\rho}$ ~~   &0 ~&2~ &2~ &1 ~&3~ &$\frac{7}{2}^+$& ~~&~&~&~&~&~&\\
\hline\hline
\end{tabular}
\end{table*}

In the present work, we will further investigate the probable assignments of the new $\Xi_{b}(6327)^{0}$ and $\Xi_{b}(6333)^{0}$ states. Furthermore, considering the powerful detecting ability of LHCb, etc., more and more $1D$ $\Xi_b$ and $\Xi'_b$ baryons are excepted to be observed in the near future. Thus, it is necessary for us to carry out a systematical study of the strong decay properties of the $1D$ $\Xi_b$ and $\Xi'_b$ baryons including both the $\rho$- and $\lambda$-mode excitations. By analyzing the decay properties of the $1D$ $\Xi_b$ and $\Xi'_b$ baryons, we will suggest ideal decay channels to establish missing states in the follow-up experiments. It should be mentioned that proper consideration of the heavy quark symmetry is necessary for the bottom baryons, wherein the states can more favor the $j$-$j$ coupling scheme~\cite{Cheng:2015iom,Wang:2018fjm,Roberts:2007ni}. Hence, we study the strong decays of the $\rho$- and $\lambda$-mode $1D$ $\Xi_b$ and $\Xi'_b$ baryons within the $j$-$j$ scheme. With this coupling scheme, the classifications of the $1D$ $\Xi_b$ and $\Xi'_b$ states investigated in this work are listed in Table~\ref{table-I}.

The paper is organized as follows. In Sec.~\ref{model}, we give a brief review of our theoretical method and the relationship between the $j$-$j$ coupling scheme and the $L$-$S$ coupling scheme. In Sec.~\ref{results}, we investigate the strong decay properties of the $\rho$- and $\lambda$-mode $1D$ $\Xi_b$ and $\Xi'_b$ baryons within the $j$-$j$ scheme, where we attempt to decode the properties of the newly observed $\Xi_{b}(6327)^{0}$ and $\Xi_{b}(6333)^{0}$ states.  Finally, we present a short summarization in Sec.~\ref{summary}.

\section{Chiral quark model}\label{model}

In this work, we systematically investigate the strong decay properties of both the $\rho$- and $\lambda$-modes $1D$ $\Xi_b$ and $\Xi'_b$ baryons within the chiral quark model~\cite{Manohar:1983md,Zhong:2007gp,Zhong:2008kd}. In the chiral quark model, the effective low energy
quark-pseudoscalar-meson coupling in the SU(3) flavor basis at tree level is given by~\cite{Manohar:1983md}
 \begin{eqnarray}\label{STcoup}
H_m=\sum_j
\frac{1}{f_m}\bar{\psi}_j\gamma^{j}_{\mu}\gamma^{j}_{5}\psi_j\vec{\tau}\cdot\partial^{\mu}\phi_m,
\end{eqnarray}
where $f_m$ represents the pseudoscalar meson decay constant and $\psi_j$ denotes the $j$-th quark field in a baryon. $\phi_m$ stands for the
pseudoscalar meson octet and reads
 \begin{eqnarray}
\phi_m=
\begin{pmatrix}
\frac{1}{\sqrt{2}}\pi^0+\frac{1}{\sqrt{6}}\eta & \pi^+ & K^+ \cr
\pi^- & -\frac{1}{\sqrt{2}}\pi^0+\frac{1}{\sqrt{6}}\eta & K^0 \cr
K^- & \bar{K}^0 & -\sqrt{\frac{2}{3}\eta}
\end{pmatrix}.
\end{eqnarray}
Considering the harmonic oscillator spatial wave
function of baryons in this work being nonrelativistic form, the quark-pseudoscalar-meson coupling is adopt the nonrelativistic form as well and      is described by~\cite{Li:1994cy,Li:1997gd,Zhao:2002id}
\begin{eqnarray}\label{non-relativistic-expansST}
H^{nr}_{m}=\sum_j\Big\{\frac{\omega_m}{E_f+M_f}\vsig_j\cdot\textbf{P}_f+ \frac{\omega_m}{E_i+M_i}\vsig_j \cdot\textbf{P}_i\\  \nonumber
-\vsig_j \cdot \textbf{q} +\frac{\omega_m}{2\mu_q}\vsig_j\cdot\textbf{p}'_j\Big\}I_j \phi_m.
\end{eqnarray}
Here, $\omega_m$ and $\textbf{q}$ correspond to the energy and three-vector momentum
of the meson, respectively; $E_{i(f)}$, $M_{i(f)}$ and $\textbf{P}_{i(f)}$ represent the energy, mass and three-vector momentum of the initial (final) baryon; The $\vsig_j$ and $\mu_q$ denote the Pauli spin vector and
the reduced mass of the $j$-th quark in the initial and final
baryons, respectively. $\textbf{p}'_j=\textbf{p}_j-(m_j/M) \textbf{P}_{c.m.}$ is the
internal momentum of the $j$-th quark in the baryon rest frame and $I_j$ is the isospin operator associated with the pseudoscalar meson; $\varphi_m=e^{(-)i\textbf{q}\cdot\textbf{r}_j}$ for absorbing (emitting) a meson.

According to the non-relativistic operator of quark-pseudoscalar-meson coupling, the partial decay amplitudes $M_{J_{iz},J_{fz}}$ of a light pseudoscalar meson emission in a baryon strong decays can be worked out. Here, $J_{iz}$ and $J_{fz}$ stand for the third components of the total angular momenta of the initial and final baryons, respectively. Then, the strong decay width can be calculated by
\begin{equation}
\Gamma=\left(\frac{\delta}{f_m}\right)^2\frac{(E_f +M_f)|q|}{4\pi
	M_i}\frac{1}{2J_i+1}\sum_{J_{iz}J_{fz}}^{}|M_{J_{iz},J_{fz}}|^2,
\end{equation}
where $\delta$ is a global parameter accounting for the strength of the quark-meson couplings,
which has been determined by experimental data in works~\cite{Zhong:2007gp,Zhong:2008kd}. Here, we fix its value the
same as that in Refs.~\cite{Zhong:2007gp,Zhong:2008kd}, i.e. $\delta=0.557$ MeV.

\begin{figure}[h]
\centering
	\centering \epsfxsize=4.0 cm \epsfbox{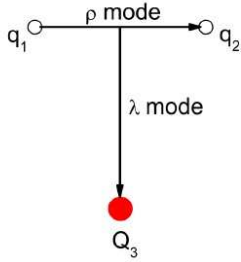}
\caption{The $\rho$- and $\lambda$-mode excitations in a single bottom baryon. Here $\rho$ and $\lambda$ are the Jacobi coordinates defined
	as $\rho=\frac{1}{\sqrt{2}}(\mathbf{r}_1-\mathbf{r}_2)$ and
	$\lambda=\frac{1}{\sqrt{6}}(\mathbf{r}_1+\mathbf{r}_2-2\mathbf{r}_3)$,
	respectively. $q_1$ and $q_2$ stand for the light quarks($u$, $s$), and $Q_3$
	stands for the bottom quark($b$). }\label{fig-str}
\end{figure}

For the single bottom baryons, decoding their inner structure within the $j$-$j$ coupling scheme is considered preferable to that within the $L$-$S$ coupling scheme (listed in Table~\ref{table1}). In the heavy quark symmetry limit~\cite{Cheng:2015iom}, the states within the the $j$-$j$ coupling scheme can be expressed as linear combinations of the states within the $L$-$S$ coupling scheme via the following relationship~\cite{Roberts:2007ni}:
\begin{equation}\label{jjls}
\begin{split}
\big{|}[[(l_{\rho}l_{ \lambda })_{L}s_{\rho}]_{j}s_{Q}]_J\big{\rangle}=
(-1)^{L+s_{\rho}+1/2+J}
 \sqrt{2j+1} \sum_{S}\sqrt{2S+1}\\
 \left(\begin{array}{ccc}
 L & s_{\rho} & j\\
 s_{Q} & J & S \\
 \end{array} \right)
\big{|}[(l_{\rho}l_{\lambda})_{L}(s_{\rho}s_{Q})_{S}]_{J^P}\big{\rangle}.
\end{split}
\end{equation}
In the expression, $l_{\rho}$ and $l_{\lambda}$ correspond to the quantum numbers of the orbital angular momenta for the $\rho$-mode and $\lambda$-mode (see Fig.~\ref{fig-str}) oscillators, respectively. $L$ ($=|l_{\rho}-l_{\lambda}|,\cdot\cdot\cdot ,l_{\rho}+l_{\lambda}$) corresponds to the quantum number of the total orbital angular momentum. $s_{\rho}$ is the quantum numbers of the total spin of the two light quarks and $s_{Q}$ is the spin of the heavy quark.  $S$ ($=|s_{\rho}-s_{Q}|,\cdot\cdot\cdot ,s_{\rho}+s_{Q}$) denotes the quantum number of the total spin angular momentum.

 \begin{table}
	\centering
	\caption{The classifications of the $1D$-wave $\Xi_b$ and $\Xi_b'$ states within the $L$-$S$ coupling scheme.
$\Xi_b$ and $\Xi_b'$ stand for the bottom baryons belonging to flavor antitriplet $\mathbf{\bar{3}}_F$ and flavor sextet $\mathbf{6}_F$, respectively.}
	\label{table1}
	\begin{tabular}{ccccccccc}
		\hline
		\hline
\multirow{1}{*}{Notation}  & \multicolumn{1}{c}{$l_\lambda$} & \multicolumn{1}{c}{$l_\rho$} & \multicolumn{1}{c}{$L$} & \multicolumn{1}{c}{$s_{\rho}$}& \multicolumn{1}{c}{$s_{Q}$}& \multicolumn{1}{c}{$S$} & \multicolumn{1}{c}{$J^P$} & \multicolumn{1}{c}{$\rm Wave\ function$}\\ \hline
${\ket{\Xi_b{^2}D_{\lambda\lambda}\frac{3}{2}^{+}}}$
		 &$0$ &$2$ &$2$&$0$ &$\frac{1}{2}$ &$\frac{1}{2}$  &$\frac{3}{2}^+$ &$^2\Psi^{\lambda\lambda}_{2L_z}\chi^{\rho}_{s_z}\phi_{\Xi_b}$ \\
${\ket{\Xi_b{^2}D_{\lambda\lambda}\frac{5}{2}^{+}}}$&$2$&$0$ &$2$ &$0$ &$\frac{1}{2}$&$\frac{1}{2}$  &$\frac{5}{2}^+$ &\\
		\hline
${\ket{\Xi_b{^2}D_{\rho\rho}\frac{3}{2}^{+}}}$
		 &$0 $&$2$ &$2$&$0$ &$\frac{1}{2}$ &$\frac{1}{2}$  &$\frac{3}{2}^+$ &$^2\Psi^{\rho\rho}_{2L_z}\chi^{\rho}_{s_z}\phi_{\Xi_b}$\\
${\ket{\Xi_b{^2}D_{\rho\rho}\frac{5}{2}^{+}}}$   &$0 $&$2$ &$2$ &$0$ &$\frac{1}{2}$&$\frac{1}{2}$  &$\frac{5}{2}^+$ &\\
		\hline
${\ket{\Xi_b^{'}{^2}D_{\lambda\lambda}\frac{3}{2}^{+}}}$
		&$2 $&$0$ &$2$&$1$ &$\frac{1}{2}$ &$\frac{1}{2}$  &$\frac{3}{2}^+$ & $^2\Psi^{\lambda\lambda}_{2L_z}\chi^{\lambda}_{s_z}\phi_{\Xi^{'}_b}$\\
${\ket{\Xi_b^{'}{^2}D_{\lambda\lambda}\frac{5}{2}^{+}}}$
		&$2 $&$0$ &$2$ &$1$ &$\frac{1}{2}$&$\frac{1}{2}$  &$\frac{5}{2}^+$ &\\
		\hline
${\ket{\Xi_b^{'}{^4}D_{\lambda\lambda}\frac{1}{2}^{+}}}$  &$2 $&$0$ &$2$&$1$ &$\frac{1}{2}$ &$\frac{3}{2}$  &$\frac{1}{2}^+$  & $^2\Psi^{\lambda\lambda}_{2L_z}\chi^S_{s_z}\phi_{\Xi^{'}_b}$ \\
${\ket{\Xi_b^{'}{^4}D_{\lambda\lambda}\frac{3}{2}^{+}}}$   &$2 $&$0$ &$2$ &$1$ &$\frac{1}{2}$&$\frac{3}{2}$  &$\frac{3}{2}^+$ &	\\
${\ket{\Xi_b^{'}{^4}D_{\lambda\lambda}\frac{5}{2}^{+}}}$   &$2 $&$0$ &$2$ &$1$ &$\frac{1}{2}$&$\frac{3}{2}$  &$\frac{5}{2}^+$ &\\
${\ket{\Xi_b^{'}{^4}D_{\lambda\lambda}\frac{7}{2}^{+}}}$  &$2 $&$0$ &$2$ &$1$ &$\frac{1}{2}$&$\frac{3}{2}$  &$\frac{7}{2}^+$ &	\\
		\hline
${\ket{\Xi_b^{'}{^2}D_{\rho\rho}\frac{3}{2}^{+}}}$
		 &$0 $&$2$ &$2$ &$1$ &$\frac{1}{2}$&$\frac{1}{2}$  &$\frac{3}{2}^+$ &$^2\Psi^{\rho\rho}_{2L_z}\chi^{\lambda}_{s_z}\phi_{\Xi^{'}_b}$\\
${\ket{\Xi^{'}_b{^2}D_{\rho\rho}\frac{5}{2}^{+}}}$   &$0 $&$2$ &$2$ &$1$ &$\frac{1}{2}$&$\frac{1}{2}$  &$\frac{5}{2}^+$ &\\
		\hline
${\ket{\Xi_b^{'}{^4}D_{\rho\rho}\frac{1}{2}^{+}}}$   &$0 $&$2$ &$2$ &$1$ &$\frac{1}{2}$&$\frac{3}{2}$  &$\frac{1}{2}^+$  & $^2\Psi^{\rho\rho}_{2L_z}\chi^S_{s_z}\phi_{\Xi^{'}_b}$ \\
		${\ket{\Xi_b^{'}{^4}D_{\rho\rho}\frac{3}{2}^{+}}}$   &$2 $&$0$ &$2$ &$1$ &$\frac{1}{2}$&$\frac{3}{2}$  &$\frac{3}{2}^+$ & 	\\
		${\ket{\Xi_b^{'}{^4}D_{\rho\rho}\frac{5}{2}^{+}}}$   &$2 $&$0$ &$2$ &$1$ &$\frac{1}{2}$&$\frac{3}{2}$  &$\frac{5}{2}^+$ &\\
		${\ket{\Xi_b^{'}{^4}D_{\rho\rho}\frac{7}{2}^{+}}}$   &$2 $&$0$ &$2$ &$1$ &$\frac{1}{2}$&$\frac{3}{2}$  &$\frac{7}{2}^+$ &	\\	
	  \hline
		\hline
		
	\end{tabular}
\end{table}

In the calculation, the standard quark model parameters are adopted. Namely, we set $ m_{u}=m_{d}=330$ MeV, $m_{s}=450 $ MeV
and $ m_{b}=5000 $ MeV for the constituent quark masses. The decay
constants for $\pi$ and $K$ mesons are taken as $ f_{\pi}=132 $ MeV and $f_{K}=160$ MeV, respectively. The masses of the initial baryons($1D$ $\Xi_b$ and $\Xi'_b$ baryons) are estimated based on the various theoretical predictions listed in Table~\ref{table-I}. The masses of the final $S$-wave ground mesons
and baryons used in the calculations are adopted from the Particle Data Group~\cite{ParticleDataGroup:2020ssz}, and those of the final $P$-wave single-bottom baryons are taken the predictions in Ref.~\cite{Ebert:2011kk}.
The spatial wave function $^{N}\Psi_{LL_z}$ of the baryons is taken the form of non-relativistic harmonic oscillator spatial-wave function. The harmonic oscillator parameter $\alpha_{\rho}$ for $us/ds$ system is taken as $\alpha_{\rho}=420$ MeV~\cite{Yao:2018jmc,Wang:2017kfr}, and another harmonic oscillator parameter $\alpha_{\lambda}$ is estimated by~\cite{Zhong:2007gp,Yao:2018jmc,Wang:2017kfr}
\begin{eqnarray}
\alpha^2_{\lambda}=\sqrt{\frac{3m_{Q}}{2m_q+m_{Q}}}\alpha^2_{\rho}.
\end{eqnarray}
We notice that in the simplified case the ratio between oscillator frequencies $\omega_{\lambda}$ and $\omega_{\rho}$ reads
\begin{eqnarray}
\frac{\omega_{\lambda}}{\omega_{\rho}}=\sqrt{\frac{1}{3}+\frac{2m_q}{3m_q}}<1.
\end{eqnarray}
This expression indicates the excitation energy of the $\lambda$-mode is smaller than that of the $\rho$-mode. Thus, the $\rho$-excitation modes are heavier than the $\lambda$-excitation modes for the $1D$ $\Xi_b$ and $\Xi'_b$ baryons.  The realistic potential is much more complicated, while the general feature should be similar.

\section{Calculations and Results }\label{results}

We conduct a systematic investigation of strong decays of $\rho$- and $\lambda$-modes $1D$ $\Xi_{b}$ and $\Xi^{'}_{b}$ within the $j$-$j$ coupling scheme in the framework of the chiral quark model, emphatically explaining the two newly discovered
states by LHCb Collaboration~\cite{LHCb:2021ssn} and giving predictions of other missing $1D$-wave states. Our theoretical results are presented as follows.

\subsection{ $ \Xi_{b} $ states }

The flavor wave functions of the $\Xi_b$ baryons belonging to flavor antitriplet $\mathbf{\bar{3}}_F$, $\phi_{\Xi_b}$, are antisymmetric, thus, their spin-spatial wave functions must be antisymmetric as well. Hence, according to the symmetry, there are two $\lambda$-mode $1D$ $\Xi_b$ baryons, $\Xi_b|J^P=\frac{3}{2}^+,2\rangle_{\lambda}$ and $\Xi_b|J^P=\frac{5}{2}^+,2\rangle_{\lambda}$, and two $\rho$-mode $1D$ $\Xi_b$ baryons, $\Xi_b|J^P=\frac{3}{2}^+,2\rangle_{\rho}$ and $\Xi_b|J^P=\frac{5}{2}^+,2\rangle_{\rho}$. The predicted masses of the two $\lambda$-mode $1D$ $\Xi_b$ baryons are about 6.3-6.4 GeV (see Table~\ref{table-I}), which are consistent with the measured masses of the newly observed $\Xi_{b}(6327)^{0}$ and $\Xi_{b}(6333)^{0}$ states at LHCb~\cite{LHCb:2021ssn}. As the possible assignments, it is crucial to investigate the decay behaviors of the two $\lambda$-mode $1D$ $\Xi_b$ baryons. For completeness, we present the prediction of the strong decays of the other two $\rho$-mode $1D$ $\Xi_b$ baryons, and hope to provide some valuable reference for the future experiment exploring.

For the $\lambda(\rho)$-mode $1D$ $\Xi_b$ baryons, there is one-to-one correspondence between the $j$-$j$ coupling scheme and the $L$-$S$ coupling scheme, namely, $|J^{P}=\frac{3}{2}^+,2\rangle_{\lambda(\rho)}=|{^2}D_{\lambda\lambda(\rho\rho)}\frac{3}{2}^{+}\rangle$ and $|J^{P}=\frac{5}{2}^+,2\rangle_{\lambda(\rho)}=|{^2}D_{\lambda\lambda(\rho\rho)}\frac{5}{2}^{+}\rangle$.

%In the $\Xi_{b}$ family, there are two $\lambda$-mode 1D-wave excitations, ${\ket{\Xi_b{^2}D_{\lambda\lambda}\frac{3}{2}^{+}}}$ and 	${\ket{\Xi_b{^2}D_{\lambda\lambda}\frac{5}{2}^{+}}}$, and two $ \rho $-mode 1D-wave excitations, ${\ket{\Xi_b{^2}D_{\rho\rho}\frac{3}{2}^{+}}}$ and 	${\ket{\Xi_b{^2}D_{\rho\rho}\frac{5}{2}^{+}}}$, expressing by the following relation with the j-j coupling scheme. The masses of  $ \lambda $-mode 1D-wave $ \Xi_{b}  $ excitations are 6320-6420 MeV within various quark model predictions, which are in good agreement with the masses of the newly observed  $\Xi^{0}_{b}(6327)$ and $\Xi^{0}_{b}(6333)$ states. On account of the excited energy of the $ \rho $-mode is higher than that of the  $ \lambda $-mode excitations, so we take a wide range of masses. Our results have been shown in FIG. 2 and FIG. 3.

\subsubsection{$\lambda $-mode excitations}

\begin{figure}[h]
	\centering \epsfxsize=8.0 cm \epsfbox{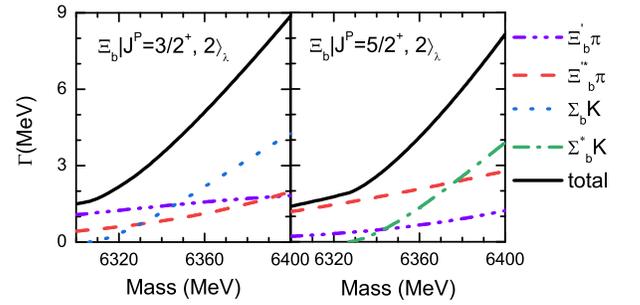}
	\caption{Partial and total strong decay widths of the $\lambda$-mode $1D$ $\Xi_b$ states as a function of the masses. The solid curves stand for the total widths. Some decay channels are not shown in the figure for their small partial decay widths.
	}\label{fig-LL}
\end{figure}

Considering the uncertainties of the predicted masses of $\Xi_b|J^P=\frac{3}{2}^+,2\rangle_{\lambda}$ and $\Xi_b|J^P=\frac{5}{2}^+,2\rangle_{\lambda}$, we plot the decay width as a function of the mass in the range of $M=(6300-6400)$ MeV in Fig.~\ref{fig-LL}. From the figure, it is found that the two $\lambda$-mode $1D$ $\Xi_b$ baryons both are narrow states with
a width of several MeV. It's important to note that the partial decay widths of $\Gamma[\Xi_b|J^P=\frac{3}{2}^+,2\rangle_{\lambda}\rightarrow\Sigma_b K]$ and $\Gamma[\Xi_b|J^P=\frac{5}{2}^+,2\rangle_{\lambda}\rightarrow\Sigma^*_b K]$ increase rapidly with the masses increasing(see Fig.~\ref{fig-LL}). Similar results were also obtained in the previous works~\cite{Chen:2019ywy,Yao:2018jmc}.

The $J^P=3/2^+$ state $\Xi_b|J^P=\frac{3}{2}^+,2\rangle_{\lambda}$ mainly decays into the $\Sigma_b K$, $\Xi'_b\pi$ and $\Xi^*_b\pi$ channels.
The $\Xi_b|J^P=\frac{3}{2}^+,2\rangle_{\lambda}$ is a good assignment of the newly observed $\Xi_b(6327)^0$ in the $\Lambda_b^0K^-\pi^+$
final state at LHCb~\cite{LHCb:2021ssn}, since it is dominantly contributed by the intermediate channel $\Sigma_b^{+}K^-$. With the measured mass $M=6327$ MeV of $\Xi_b(6327)^0$, our predicted decay properties of $\Xi_b|J^P=\frac{3}{2}^+,2\rangle_{\lambda}$ have
been shown in Table~\ref{Table3}. It is seen that the total decay width
\begin{eqnarray}
\Gamma_{\text{Total}}\simeq2.56 \ \ \mathrm{MeV},
\end{eqnarray}
is consistent with the observations. The branching fraction for the $\Sigma_bK$ channel is
\begin{eqnarray}
\frac{\Gamma[\Xi_b|J^P=\frac{3}{2}^+,2\rangle_{\lambda}\rightarrow\Sigma_b K]}{\Gamma_{\text{Total}}}\sim 23\%.
\end{eqnarray}
In addition, we get that
\begin{eqnarray}
\frac{\Gamma[\Xi_b|J^P=\frac{3}{2}^+,2\rangle_{\lambda}\rightarrow\Xi'_b \pi]}{\Gamma_{\text{Total}}}\sim 51\%,\\
\frac{\Gamma[\Xi_b|J^P=\frac{3}{2}^+,2\rangle_{\lambda}\rightarrow\Xi^{'*}_b \pi]}{\Gamma_{\text{Total}}}\sim 26\%.
\end{eqnarray}
If the newly observed $\Xi_b(6327)^0$ state corresponds to $\Xi_b|J^P=\frac{3}{2}^+,2\rangle_{\lambda}$ indeed, besides the $\Sigma_bK$ channel, the $\Xi'_b \pi$ and $\Xi^{'*}_b \pi$ may be another two interesting channels for the observation of $\Xi_b(6327)$ in future experiments.
The $\Xi_b(6327)$ resonance should be observed in the $\Xi_b \pi\pi$ final state as well.

\begin{table}
\caption{The partial decay widths(MeV) of $\Xi_b(6327)^0$ and $\Xi_b(6333)^0$ assigned as $\lambda$-mode $1D$ $\Xi_b$ baryons $\Xi_b\Ket{J^{P}=\frac{3}{2}^+,2}_{\lambda}$ and $\Xi_b\Ket{J^{P}=\frac{5}{2}^+,2}_{\lambda}$, respectively.}\label{Table3}
	\centering
	\begin{tabular}{c|ccccccc}\hline\hline
		\multirow{2}{*}{Decay width} & $\underline{~~~~~~\Xi_{b}|J^{P}=\frac{3}{2}^+,2 \rangle_{\lambda}~~~~~~}$ & $\underline{~~~~~~\Xi_{b}|J^{P}=\frac{5}{2}^+,2\rangle_{\lambda}~~~~~~}$\\
		&$\Xi_b(6327)^0$ &$\Xi_b(6333)^0$\\ \hline
        $ \Gamma[\Sigma_bK] $&0.59&0.00\\
		$ \Gamma[\Sigma^{*}_bK] $&-&0.11\\
		$ \Gamma[\Xi^{'}_{b}\pi] $&1.30&0.41\\
		$ \Gamma[\Xi^{'*}_{b}\pi] $&0.67&1.64\\
		$\Gamma_{\text{Total}}$ & 2.56&2.16\\
		Expt.&$ < $2.20(2.56)&$ < $1.55(1.85)\\
	\hline\hline
	\end{tabular}
\end{table}

For the $J^P=5/2^+$ state $\Xi_b|J^P=\frac{5}{2}^+,2\rangle_{\lambda}$ (see Fig.~\ref{fig-LL}), the mainly decay channels are the $\Xi'_b\pi$, $\Sigma^*_b K$ and $\Xi^*_b\pi$ channels. Combining the natures of the newly observed state $\Xi_b(6333)^0$, we obtain that this new state may be an assignment of  $\Xi_b|J^P=\frac{5}{2}^+,2\rangle_{\lambda}$. Fixing the mass of $\Xi_b|J^P=\frac{5}{2}^+,2\rangle_{\lambda}$ at $M=6333$ MeV, we collect its decay properties in Table~\ref{Table3} as well. It is found that
the total decay width
\begin{eqnarray}
\Gamma_{\text{Total}}\simeq2.16 \ \ \mathrm{MeV},
\end{eqnarray}
is close to the upper limit of the observed one. The branching fractions for the main decay channels are predicted to be

\begin{eqnarray}
\frac{\Gamma[\Xi_b|J^P=\frac{5}{2}^+,2\rangle_{\lambda}\rightarrow\Sigma^*_b K]}{\Gamma_{\text{Total}}}\sim 5\%,\\
\frac{\Gamma[\Xi_b|J^P=\frac{5}{2}^+,2\rangle_{\lambda}\rightarrow\Xi'_b \pi]}{\Gamma_{\text{Total}}}\sim 19\%,\\
\frac{\Gamma[\Xi_b|J^P=\frac{5}{2}^+,2\rangle_{\lambda}\rightarrow\Xi^{'*}_b \pi]}{\Gamma_{\text{Total}}}\sim 76\%.
\end{eqnarray}
Thus, basing on our calculations, these strong decay processes may be measured due to their significant branching fractions.
To confirm the $\Xi_b(6333)$ resonance, the $\Xi_b \pi\pi$ final state is worth observing in experiments.

\subsubsection{$\rho $-mode excitations}

\begin{figure}[h]
	\centering \epsfxsize=8.0 cm \epsfbox{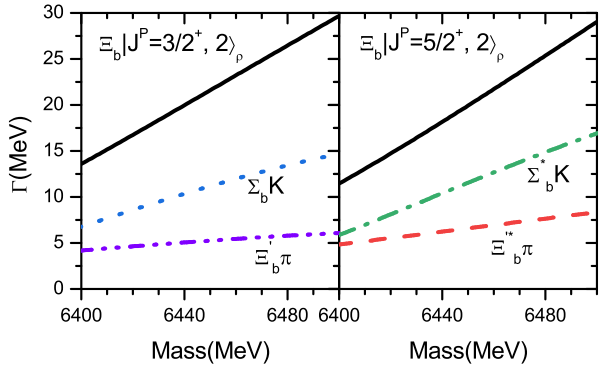}
	\caption{Partial and total strong decay widths of the $\rho$-mode $1D$ $\Xi_b$ states as a function of the masses. The solid curves stand for the total widths. Some decay channels are not shown in the figure for their small partial decay widths.}\label{fig-RR}
\end{figure}

There are some qualitative discussions for the masses of $\rho$-mode $1D$ $\Xi_b$ states~\cite{Narodetskii:2008pn}, and pointed out that the masses of $\rho$-mode excitations were about 100 MeV heavier than those of $ \lambda$-mode excitations. Hence, the masses of the two $\rho$-mode $1D$ $\Xi_b$ states, $\Xi_b|J^P=\frac{3}{2}^+,2\rangle_{\rho}$ and $\Xi_b|J^P=\frac{5}{2}^+,2\rangle_{\rho}$, may vary in the range of $M=(6400-6500)$ MeV, and we are most likely to exclude the two $\rho$-mode states as assignments of $\Xi_b(6327)^0$ and $\Xi_b(6333)^0$ based on the masses. We calculate the decay properties of the two $\rho$-mode $1D$ $\Xi_b$ states as a function of the mass within the possible range allowed, as shown in Fig.~\ref{fig-RR}. The total decay widths of the two $\rho$-mode $1D$ $\Xi_b$ states are about $\Gamma\simeq(12-30)$ MeV within the mass range what we considered.

Meanwhile, we notice that $\Xi_b|J^P=\frac{3}{2}^+,2\rangle_{\rho}$ decays mainly through the $\Sigma_bK$ channel, and the predicted branching ratio is
\begin{eqnarray}
\frac{\Gamma[\Xi_b|J^P=\frac{3}{2}^+,2\rangle_{\rho}\rightarrow\Sigma_b K]}{\Gamma_{\text{Total}}}\sim 51\%.
\end{eqnarray}
For the $\Xi_b|J^P=\frac{5}{2}^+,2\rangle_{\rho}$ state, the main decay channel is $\Sigma^*_bK$ and the corresponding branching fraction is
\begin{eqnarray}
\frac{\Gamma[\Xi_b|J^P=\frac{5}{2}^+,2\rangle_{\rho}\rightarrow\Sigma^*_b K]}{\Gamma_{\text{Total}}}\sim (53-58)\%.
\end{eqnarray}

If we don't care about masses of $\Xi_b|J^P=\frac{3}{2}^+,2\rangle_{\rho}$ and $\Xi_b|J^P=\frac{5}{2}^+,2\rangle_{\rho}$, the two $\rho$-mode $1D$ $\Xi_b$ states are good candidates of the newly observed states $\Xi_b(6327)^0$ and $\Xi_b(6333)^0$, respectively, where the $\Sigma_bK$ process dominates the decay of $\Xi_b|J^P=\frac{3}{2}^+,2\rangle_{\rho}$, while the $\Sigma^*_bK$ channel dominates the decay of $\Xi_b|J^P=\frac{5}{2}^+,2\rangle_{\rho}$.
To further clarify the inner structures of the two newly observed states and verify our predictions, more experimental observations are needed.

\subsection{ $ \Xi^{'}_{b} $ states }

The flavor wave functions of the $\Xi'_b$ baryons belonging to sextet $\mathbf{6}_F$, $\phi_{\Xi'_b}$, are symmetric, thus, their spin-spatial wave functions must be symmetric as well. Based on the symmetry, there are six $\lambda$-mode $1D$ $\Xi'_b$ baryons and six $\rho$-mode $1D$ $\Xi'_b$ baryons, as listed in Table~~\ref{table-I}. According to the theoretical predictions by various quark model, the masses of $\lambda$-mode $1D$ $\Xi^{'}_{b}$ baryons vary in the region of $M=(6380-6480)$ MeV. Considering the mass of the $\rho$-mode excitation being $\sim100$ MeV heavier than the $\lambda$-mode excitation, the $\rho$-mode $1D$  $\Xi^{'}_{b}$ baryons may be in the range of $M=(6480-6580)$ MeV. From the point of view of mass, the possibility of the newly observed $\Xi^{0}_{b}(6327)$ and $\Xi^{0}_{b}(6333)$ states as the $\Xi^{'}_{b}$ state may be excluded.

From relationship given in Eq.~(\ref{jjls}),
the $1D$ $\Xi'_b$ baryon states in the $j$-$j$ coupling scheme can be expressed with the linear combination of the configurations in the $L$-$S$ coupling scheme:
\begin{equation}
\Ket{J^{P}=\frac{1}{2}^+,1}_{\lambda(\rho)}=\Ket{{^4}D_{\lambda\lambda(\rho\rho)}\frac{1}{2}^{+}},
\end{equation}
\begin{equation}
\Ket{J^{P}=\frac{3}{2}^+,1}_{\lambda(\rho)}=\sqrt{\frac{1}{2}}\Ket{{^4}D_{\lambda\lambda(\rho\rho)}\frac{3}{2}^{+}}-\sqrt{\frac{1}{2}}\Ket{{^2}D_{\lambda\lambda(\rho\rho)}\frac{3}{2}^{+}},
\end{equation}
\begin{equation}
\Ket{J^{P}=\frac{3}{2}^+,2}_{\lambda(\rho)}=\sqrt{\frac{1}{2}}\Ket{{^4}D_{\lambda\lambda(\rho\rho)}\frac{3}{2}^{+}}+\sqrt{\frac{1}{2}}\Ket{{^2}D_{\lambda\lambda(\rho\rho)}\frac{3}{2}^{+}},
\end{equation}
\begin{equation}
\Ket{J^{P}=\frac{5}{2}^+,2}_{\lambda(\rho)}=\frac{\sqrt{7}}{3}\Ket{{^4}D_{\lambda\lambda(\rho\rho)}\frac{5}{2}^{+}}-\frac{\sqrt{2}}{3}\Ket{{^2}D_{\lambda\lambda(\rho\rho)}\frac{5}{2}^{+}},
\end{equation}
\begin{equation}
\Ket{J^{P}=\frac{5}{2}^+,3}_{\lambda(\rho)}=\frac{\sqrt{2}}{3}\Ket{{^4}D_{\lambda\lambda(\rho\rho)}\frac{5}{2}^{+}}+\frac{\sqrt{7}}{3}\Ket{{^2}D_{\lambda\lambda(\rho\rho)}\frac{5}{2}^{+}},
\end{equation}
\begin{equation}
\Ket{J^{P}=\frac{7}{2}^+,3}_{\lambda(\rho)}=\Ket{{^4}D_{\lambda\lambda(\rho\rho)}\frac{7}{2}^{+}}.
\end{equation}
In the following, we present our theoretical predictions of the $1D$ $\Xi'_b$ baryons within the $j$-$j$ coupling scheme.

\subsubsection{$\lambda $-mode excitations}

Firstly, we fix the masses of the $\lambda$-mode $1D$ $\Xi^{'}_{b}$ states at the predictions within the nonrelativistic quark-diquark picture in Ref.~\cite{Ebert:2011kk}, and collect the decay properties in Table~\ref{Table4}.

\begin{table*}
\caption{The decay properties of the $\lambda$- and $\rho-$ modes $1D$ $\Xi^{'}_{b}$ states. $\Gamma_{\text{Total}}$ stands for the total decay width. The unit of the mass and width is MeV. In this work the $1P$-wave states $|\Lambda^{2}_{b}P_{\lambda}\frac{1}{2}^{-}\rangle$ and $|\Lambda^{2}_{b}P_{\lambda}\frac{3}{2}^{-}\rangle$ are assigned to $\Lambda_b(5912)$ and $\Lambda_b(5920)$, respectively.
The masses for the unestablished $1P$-wave $\Xi_b$ and $\Xi_b'$ states are taken the predictions in Ref.~\cite{Ebert:2011kk}. }\label{Table4}
	\centering
	\begin{tabular}{cccccccc}\hline\hline
		\multirow{2}{*}{Decay width} & $\underline{~~\Xi'_{b}|J^{P}=\frac{1}{2}^+,1 \rangle_{\lambda}~~}$ & $\underline{~~\Xi'_{b}|J^{P}=\frac{3}{2}^+,1\rangle_{\lambda}~~}$ & $\underline{~~\Xi'_{b}|J^{P}=\frac{3}{2}^+,2 \rangle_{\lambda}~~}$& $\underline{~~\Xi'_{b}|J^{P}=\frac{5}{2}^+,2 \rangle_{\lambda~~}}$& $\underline{~~\Xi'_{b}|J^{P}=\frac{5}{2}^+,3 \rangle_{\lambda}~~}$& $\underline{~~\Xi'_{b}|J^{P}=\frac{7}{2}^+,3 \rangle_{\lambda}~~}$\\
	                 &$M=6447$ &$M=6459$  &$M=6431$ &$M=6432$  &$M=6420$ &$M=6414$  \\ \hline
$\Gamma$[$\Xi_{b}$$\pi$] & 2.21&1.96&-&-&9.60&9.11\\
	 $\Gamma$[$\Xi^{'}_{b}$$\pi$]&1.30&0.32&2.89&1.27&1.23&0.64\\
	 $\Gamma$[$\Xi^{'*}_{b}$$\pi$]&0.64&1.62&2.28&4.06&2.09&0.64\\
	  $\Gamma$[$\Lambda_{b}$K]&2.41&2.10&-&-&5.77&5.43\\
	   $\Gamma$[$\Sigma_{b}$K]&4.08&1.07&8.37&0.47&0.38&0.17\\
	  $\Gamma$[$\Sigma^{*}_{b}$K] &1.76&4.86&1.84&6.83&0.77&0.10\\
	   $\Gamma$[$|\Lambda^{2}_{b}P_{\lambda}\frac{1}{2}^{-}\rangle$K]&3.14&3.03&-&-&-&0.05\\
	    $\Gamma$[$|\Lambda^{2}_{b}P_{\lambda}\frac{3}{2}^{-}\rangle$K]&6.24&6.24&-&-&-&-\\
	    $\Gamma$[$|\Xi^{2}_{b}P_{\lambda}\frac{1}{2}^{-}\rangle$$\pi$] &0.85&0.76&-&-&0.43&1.59\\
	     $\Gamma$[$|\Xi^{2}_{b}P_{\lambda}\frac{3}{2}^{-}\rangle$$\pi$]&1.83&1.67&-&-&0.70&2.83\\
	      $\Gamma$[$|\Xi^{'2}_{b}P_{\lambda}\frac{1}{2}^{-}\rangle$$\pi$]&0.87&0.08&0.60&0.01&0.01&0.06\\
	      $\Gamma$[$|\Xi^{'2}_{b}P_{\lambda}\frac{3}{2}^{-}\rangle$$\pi$]&0.61&0.16&1.19&0.02&0.01&0.11\\
	      $\Gamma$[$|\Xi^{'4}_{b}P_{\lambda}\frac{1}{2}^{-}\rangle$$\pi$] &0.16&0.04&0.07&0.56&0.01&0.04\\
	       $\Gamma$[$|\Xi^{'4}_{b}P_{\lambda}\frac{3}{2}^{-}\rangle$$\pi$]&0.03&0.01&0.02&0.12&-&0.01\\
	       $\Gamma$[$|\Xi^{'4}_{b}P_{\lambda}\frac{5}{2}^{-}\rangle$$\pi$] &1.17&0.08&0.13&1.03&0.01&0.07\\
	       $\Gamma_{\text{Total}}$&26.87&23.93&17.39&14.37&21.27&21.09\\
\hline\hline
\multirow{2}{*}{Decay width} & $\underline{~~\Xi'_{b}|J^{P}=\frac{1}{2}^+,1 \rangle_{\rho}~~}$ & $\underline{~~\Xi'_{b}|J^{P}=\frac{3}{2}^+,1\rangle_{\rho}~~}$ & $\underline{~~\Xi'_{b}|J^{P}=\frac{3}{2}^+,2 \rangle_{\rho}~~}$& $\underline{~~\Xi'_{b}|J^{P}=\frac{5}{2}^+,2 \rangle_{\rho}~~}$& $\underline{~~\Xi'_{b}|J^{P}=\frac{5}{2}^+,3 \rangle_{\rho}~~}$& $\underline{~~\Xi'_{b}|J^{P}=\frac{7}{2}^+,3 \rangle_{\rho}~~}$\\
	                 &$M=6547$ &$M=6559$  &$M=6531$ &$M=6532$  &$M=6520$ &$M=6514$  \\ \hline
 $\Gamma$[$\Xi_{b}$$\pi$] &11.00&10.56&-&-&10.52&10.12\\
	 $\Gamma$[$\Xi^{'}_{b}$$\pi$]&4.34&1.09&9.61&2.00&2.03&1.08\\
	 $\Gamma$[$\Xi^{'*}_{b}$$\pi$]&2.13&5.38&4.83&11.44&5.26&2.66\\
	  $\Gamma$[$\Lambda_{b}$K]&7.64&7.16&-&-&6.73&6.47\\
	   $\Gamma$[$\Sigma_{b}$K]&11.06&2.81&24.15&1.79&1.70&0.87\\
	  $\Gamma$[$\Sigma^{*}_{b}$K] &4.42&13.75&7.15&21.12& 5.31&1.96\\
	   $\Gamma$[$|\Lambda^{2}_{b}P_{\lambda}\frac{1}{2}^{-}\rangle$K]&0.49&0.58&-&-&0.03&0.02\\
	    $\Gamma$[$|\Lambda^{2}_{b}P_{\lambda}\frac{3}{2}^{-}\rangle$K]&0.87&1.94&0.08&-&0.04&0.03\\
	    $\Gamma$[$|\Xi^{2}_{b}P_{\lambda}\frac{1}{2}^{-}\rangle$$\pi$] &0.47&0.53&-&-&0.06&0.05\\
	     $\Gamma$[$|\Xi^{2}_{b}P_{\lambda}\frac{3}{2}^{-}\rangle$$\pi$]&0.83&1.87&0.10&-&0.10&0.09\\
	      $\Gamma$[$|\Xi^{'2}_{b}P_{\lambda}\frac{1}{2}^{-}\rangle$$\pi$]&0.03&0.01&0.04&-&-&-\\
	      $\Gamma$[$|\Xi^{'2}_{b}P_{\lambda}\frac{3}{2}^{-}\rangle$$\pi$]&0.05&0.02&0.08&-&-&-\\
	      $\Gamma$[$|\Xi^{'4}_{b}P_{\lambda}\frac{1}{2}^{-}\rangle$$\pi$] &0.01&-&0.01&0.05&-&-\\
	       $\Gamma$[$|\Xi^{'4}_{b}P_{\lambda}\frac{3}{2}^{-}\rangle$$\pi$]&-&-&-&0.01&-&-\\
	       $\Gamma$[$|\Xi^{'4}_{b}P_{\lambda}\frac{5}{2}^{-}\rangle$$\pi$] &0.03&0.01&0.02&0.09&-&-\\
	       $\Gamma_{\text{Total}}$&43.46&45.71&46.07&36.5&31.78&23.35\\ \hline   \hline
\end{tabular}
\end{table*}

\begin{figure*}[]
	\centering \epsfxsize=15.0 cm \epsfbox{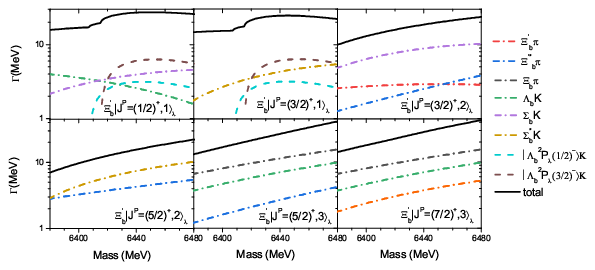}
	\caption{Partial and total strong decay widths of the $\lambda$-mode $1D$ $\Xi'_b$ states as a function of the masses. Some decay channels are not shown in the figure for their small partial decay widths.}\label{fig-LLprime}
\end{figure*}

The $J^P=1/2^+$ state $\Xi'_{b}|J^{P}=\frac{1}{2}^+, 1\rangle_{\lambda}$ has a narrow width of $\Gamma\simeq27$ MeV
and has relatively large decay rates into the $\Lambda_bK$, $\Sigma_bK$, $\Xi_b\pi$, $\Lambda_b(5912) K$ and
$ \Lambda_b(5920)K$ channels. The branching fractions for the $\Lambda_bK$, $\Sigma_bK$, $\Xi_b\pi$ channels are predicted to be
\begin{eqnarray}
\frac{\Gamma[\Xi'_b|J^P=\frac{1}{2}^+,1\rangle_{\lambda}\to\Lambda_bK]}{\Gamma_{\text{Total}}}\sim 9\%,\\
\frac{\Gamma[\Xi'_b|J^P=\frac{1}{2}^+,1\rangle_{\lambda}\to\Sigma_bK]}{\Gamma_{\text{Total}}}\sim 15\%,\\
\frac{\Gamma[\Xi'_b|J^P=\frac{1}{2}^+,1\rangle_{\lambda}\to\Xi_b\pi]}{\Gamma_{\text{Total}}}\sim 8\%.
\end{eqnarray}
The $\Lambda_bK$, $\Sigma_bK$ and $\Xi_b\pi$ channels can be used to search for the missing $\Xi'_{b}|J^{P}=\frac{1}{2}^+, 1\rangle_{\lambda}$ state.

For the $J^P=3/2^+$ state $\Xi'_b|J^{P}=\frac{3}{2}^+,1\rangle_{\lambda} $, its width is predicted to be around
$\Gamma\simeq24$ MeV. This state has large decay rates into  $ \Lambda_b(5920)K$, $\Sigma^{*}_bK$ and $\Lambda_b(5912) K$.
The branching fraction for the $\Sigma^{*}_bK$ channel can reach up to
\begin{eqnarray}
\frac{\Gamma[\Xi'_b|J^P=\frac{3}{2}^+,1\rangle_{\lambda}\to\Sigma^*_bK]}{\Gamma_{\text{Total}}}\sim 20\%.
\end{eqnarray}
The $\Xi'_b|J^{P}=\frac{3}{2}^+,1\rangle_{\lambda} $ may have a large potential to be observed in the
$\Lambda_b\pi K$ final state via the decay chain
$\Xi'_b|J^{P}=\frac{3}{2}^+,1\rangle_{\lambda}\to \Sigma^*_bK\to \Lambda_b\pi K$ at LHCb.

The other $J^P=3/2^+$ state $\Xi'_b|J^{P}=\frac{3}{2}^+,2\rangle_{\lambda}$ has a
width of $\Gamma\simeq17$ MeV, and dominantly decay into $\Sigma_bK$ channel with a branching fraction
\begin{eqnarray}
\frac{\Gamma[\Xi'_b|J^P=\frac{3}{2}^+,2\rangle_{\lambda}\rightarrow \Sigma_bK]}{\Gamma_{\text{Total}}}\sim 48\%.
\end{eqnarray}
This state may have a large potential to be observed in the
$\Lambda_b\pi K$ final state as well via the decay chain
$\Xi'_b|J^{P}=\frac{3}{2}^+,1\rangle_{\lambda}\to \Sigma_bK\to \Lambda_b\pi K$ at LHCb.

For the $J^P=5/2^+$ state $\Xi'_b|J^P=\frac{5}{2}^+,2\rangle_{\lambda}$, the width is predicted
to be $\Gamma\simeq 14$ MeV. This state mainly decays into $\Sigma^*_bK$ and $\Xi_b^{'*}\pi$.
Their branching fractions are
\begin{eqnarray}
\frac{\Gamma[\Xi'_b|J^P=\frac{5}{2}^+,2\rangle_{\lambda}\rightarrow\Sigma^*_bK]}{\Gamma_{\text{Total}}}\sim 47\%,\\
\frac{\Gamma[\Xi'_b|J^P=\frac{5}{2}^+,2\rangle_{\lambda}\rightarrow \Xi_b^{'*}\pi]}{\Gamma_{\text{Total}}}\sim 28\%.
\end{eqnarray}
The $\Xi'_b|J^P=\frac{5}{2}^+,2\rangle_{\lambda}$ may be observed in the $\Lambda_b\pi K$ or/and
$\Xi_b\pi\pi$ final states via the decay chains
$\Xi'_b|J^{P}=\frac{5}{2}^+,2\rangle_{\lambda}\to \Sigma_b^*K/\Xi_b^{'*}\pi\to \Lambda_b\pi K/\Xi_b\pi\pi$.

Both $\Xi'_b|J^P=\frac{5}{2}^+,3\rangle_{\lambda}$ and $\Xi'_b|J^P=\frac{7}{2}^+,3\rangle_{\lambda}$
have a similar width of $\Gamma\simeq21$ MeV, and dominantly decay into $\Xi_b\pi$ and $\Lambda_bK$ channels.
The branching fractions are predicted to be
\begin{eqnarray}
\frac{\Gamma[\Xi'_b|J^P=\frac{5}{2}^+(\frac{7}{2}^+),3\rangle_{\lambda}\to\Lambda_bK]}{\Gamma_{\text{Total}}}\sim 28\% (26\%),\\
\frac{\Gamma[\Xi'_b|J^P=\frac{5}{2}^+(\frac{7}{2}^+),3\rangle_{\lambda}\rightarrow \Xi_b \pi]}{\Gamma_{\text{Total}}}\sim 45\% (43\%).
\end{eqnarray}
The $\Xi_b\pi$ and $\Lambda_bK$ may be good channels to search for the $\Xi'_b|J^P=\frac{5}{2}^+,3\rangle_{\lambda}$ and $\Xi'_b|J^P=\frac{7}{2}^+,3\rangle_{\lambda}$ states. Furthermore, it is found that
the $\Xi'_b|J^P=\frac{5}{2}^+,3\rangle_{\lambda}$ state has sizeable decay rates into the $\Xi_b^{'*}\pi$ channel
with a branching fraction of
\begin{eqnarray}
\frac{\Gamma[\Xi'_b|J^P=\frac{5}{2}^+,3\rangle_{\lambda}\rightarrow \Xi^{'*}_b \pi]}{\Gamma_{\text{Total}}}\sim 10\%.
\end{eqnarray}
The $\Xi_b^{'*}\pi$ channel which can be used to distinguish $\Xi'_b|J^P=\frac{5}{2}^+,3\rangle_{\lambda}$ from $\Xi'_b|J^P=\frac{7}{2}^+,3\rangle_{\lambda}$ in future experiments.

The predicted masses of the $\lambda-$ mode $1D$ $\Xi'_b$ baryons certainly have a large uncertainty, which may bring uncertainties to the theoretical results. To investigate this effect, we plot the two-body strong decay widths of the $\lambda$-mode $1D$ $\Xi'_b$ baryons as a function of the mass in Fig.~\ref{fig-LLprime}. The sensitivities of the decay properties of these states to their masses can be clearly seen from the figure. As a whole, the $\lambda$-mode $1D$ $\Xi^{'}_{b}$ states have a fairly narrow width
of $\Gamma\simeq(13-27)$ MeV. To looking for these missing states, the $\Lambda_b K$,
$\Xi_b\pi$, $\Lambda_b K \pi$ and $\Xi_b\pi\pi$ are worth observing in future experiments.

\subsubsection{$\rho $-mode excitations}

Considering the masses of the $\rho$-mode excitations being about 100 MeV heavier than the $\lambda$-mode excitations, thus, we fix the masses of the six $\rho$-mode $1D$ $\Xi'_b$ baryons in the range of $M=(6480-6580)$ MeV, and listed the decay properties in Table~\ref{Table4} as well.

\begin{figure*}[]
	\centering \epsfxsize=15.0 cm \epsfbox{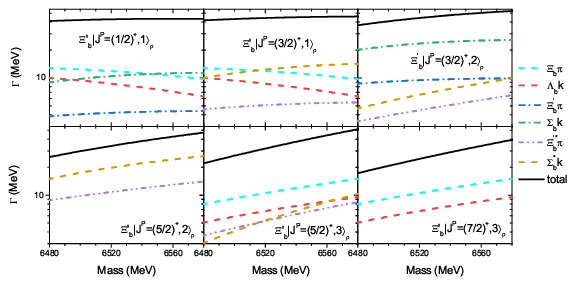}
	\caption{Partial and total strong decay widths of the $\rho$-mode $1D$ $\Xi'_b$ states as a function of the masses. Some decay channels are not shown in the figure for their small partial decay widths.}\label{fig-RRprime}
\end{figure*}

 The three $\rho$-mode states $\Xi'_b|J^P=\frac{1}{2}^+,1\rangle_{\rho}$, $\Xi'_b|J^P=\frac{3}{2}^+,1\rangle_{\rho}$
 and $\Xi'_b|J^P=\frac{3}{2}^+,2\rangle_{\rho}$ have a comparable width of $\Gamma\simeq45$ MeV. While, we notice that the main decay channels have some different among those three states. The $\Xi'_b|J^P=\frac{1}{2}^+,1\rangle_{\rho}$ mainly decays into $\Sigma_bK$, $\Lambda_bK$, and $\Xi_b\pi$ channels. Their branching fractions are predicted to be
 \begin{eqnarray}
 \frac{\Gamma[\Xi'_b|J^P=\frac{1}{2}^+,1\rangle_{\rho}\to\Sigma_bK]}{\Gamma_{\text{Total}}}\sim 25\%,\\
 \frac{\Gamma[\Xi'_b|J^P=\frac{1}{2}^+,1\rangle_{\rho}\to\Lambda_bK]}{\Gamma_{\text{Total}}}\sim 17\%,\\
 \frac{\Gamma[\Xi'_b|J^P=\frac{1}{2}^+,1\rangle_{\rho}\to\Xi_b\pi]}{\Gamma_{\text{Total}}}\sim 25\%.
 \end{eqnarray}
 Meanwhile, the $\Xi'_b|J^P=\frac{1}{2}^+,1\rangle_{\rho}$ has sizeable decay rates
 into $\Xi'_b\pi$ and $\Sigma_b^*K$, their branching fractions can reach up to $\sim10\%$.

For the $\Xi'_b|J^P=\frac{3}{2}^+,1\rangle_{\rho}$ state, the dominant decay modes are $\Sigma^*_bK$ and $\Xi_b\pi$ with branching fractions
\begin{eqnarray}
 \frac{\Gamma[\Xi'_b|J^P=\frac{3}{2}^+,1\rangle_{\rho}\rightarrow \Sigma^*_bK]}{\Gamma_{\text{total}}}\sim30\%,\\
 \frac{\Gamma[\Xi'_b|J^P=\frac{3}{2}^+,1\rangle_{\rho}\rightarrow \Xi_b\pi]}{\Gamma_{\text{total}}}\sim23\%.
 \end{eqnarray}
The decay rates of $\Xi'_b|J^P=\frac{3}{2}^+,1\rangle_{\rho}$ into $\Xi_b^{'*}\pi$ and $\Lambda_bK$ are fairly large as well. Their  branching fractions are predicted to be about $11\%$ and $15\%$, respectively.

The decay of $\Xi'_b|J^P=\frac{3}{2}^+,2\rangle_{\rho}$ is governed by $\Sigma_bK$. The branching fraction is predicted to be
\begin{eqnarray}
 \frac{\Gamma[\Xi'_b|J^P=\frac{3}{2}^+,2\rangle_{\rho}\rightarrow \Sigma_bK]}{\Gamma_{\text{total}}}\sim52\%.
 \end{eqnarray}
The rates of $\Xi'_b|J^P=\frac{3}{2}^+,2\rangle_{\rho}$ decaying into $\Xi'_b\pi$ and $\Sigma^*_bK$ are sizable and predicted to
be
\begin{eqnarray}
 \frac{\Gamma[\Xi'_b|J^P=\frac{3}{2}^+,2\rangle_{\rho}\rightarrow \Xi'_b\pi]}{\Gamma_{\text{total}}}\sim21\%, \\
 \frac{\Gamma[\Xi'_b|J^P=\frac{3}{2}^+,2\rangle_{\rho}\rightarrow \Sigma^*_bK]}{\Gamma_{\text{total}}}\sim15\%.
 \end{eqnarray}

Compared with the three states above, the total decay widths of the two $J^P=5/2^+$ states $\Xi'_b|J^P=\frac{5}{2}^+,2\rangle_{\rho}$ and $\Xi'_b|J^P=\frac{5}{2}^+,3\rangle_{\rho}$ are a little narrower with a width of $\Gamma\simeq35$ MeV. The $\Xi'_b|J^P=\frac{5}{2}^+,2\rangle_{\rho}$ state mainly decays into $\Sigma^*_bK$ and $\Xi^{'*}_b\pi$ with branching fractions
\begin{eqnarray}
 \frac{\Gamma[\Xi'_b|J^P=\frac{5}{2}^+,2\rangle_{\rho}\rightarrow \Sigma^*_bK]}{\Gamma_{\text{total}}}\sim58\%, \\
 \frac{\Gamma[\Xi'_b|J^P=\frac{5}{2}^+,2\rangle_{\rho}\rightarrow \Xi^{'*}_b\pi]}{\Gamma_{\text{total}}}\sim31\%.
 \end{eqnarray}
While, the strong decay of the state $\Xi'_b|J^P=\frac{5}{2}^+,3\rangle_{\rho}$ is dominated by the $\Xi_b\pi$ channel, and the branching ratio is
\begin{eqnarray}
 \frac{\Gamma[\Xi'_b|J^P=\frac{5}{2}^+,3\rangle_{\rho}\rightarrow \Xi_b\pi]}{\Gamma_{\text{total}}}\sim33\%,
 \end{eqnarray}
which can be used to distinguish $\Xi'_b|J^P=\frac{5}{2}^+,3\rangle_{\rho}$ from $\Xi'_b|J^P=\frac{5}{2}^+,2\rangle_{\rho}$
in future experiments. The $\Xi'_b|J^P=\frac{5}{2}^+,3\rangle_{\rho}$ also has relatively large  decay rates into the
$\Xi^{'*}_b\pi$, $\Lambda_bK$, and $\Sigma^*_bK$ channels with a comparable branching fraction of $\sim15\%$.

The $\Xi'_b|J^P=\frac{7}{2}^+,3\rangle_{\rho}$ may be the narrowest one among the six $\rho$-mode $1D$ $\Xi'_b$ baryons with a width of $\Gamma\simeq23$ MeV, and mainly decays into $\Xi_b\pi$ and $\Lambda_bK$ channel. The predicted branching fractions are
\begin{eqnarray}
 \frac{\Gamma[\Xi'_b|J^P=\frac{7}{2}^+,3\rangle_{\rho}\to \Xi_b\pi]}{\Gamma_{\text{total}}}\sim43\%.\\
 \frac{\Gamma[\Xi'_b|J^P=\frac{7}{2}^+,3\rangle_{\rho}\to \Lambda_bK]}{\Gamma_{\text{total}}}\sim28\%.
 \end{eqnarray}

Similarly, considering the uncertainties of the masses, we also plot the the variations of the partial decay widths as a function of the mass, and show in Fig.~\ref{fig-RRprime}. As a whole the $\rho$-mode $1D$ $\Xi'_b$ baryons may also have good potentials to be observed in experiments due to
their relatively narrow widths. The ideal channels for observations are $\Xi_b\pi$, $\Lambda_bK$, $\Sigma_bK$,
$\Sigma^*_bK$ and $\Xi^{'*}_b\pi$.

\section{Summary}\label{summary}

Stimulated by the newly observed bottom baryon resonances $\Xi^{0}_{b}(6327)$ and $\Xi^{0}_{b}(6333)$ at LHCb, we carry out a systematic study on the two-body strong decay behaviors of the $\rho$- and $\lambda$-mode $1D$ $\Xi_{b}$ and $\Xi^{'}_{b}$ baryons in the framework of chiral quark model within the $j$-$j$ coupling scheme. For the newly observed states $\Xi^{0}_{b}(6327)$ and $\Xi^{0}_{b}(6333)$, we give a possible theoretical interpretation. Meanwhile, we give the predictions for the strong decay properties of the missing $1D$ $\Xi_{b}$ and $\Xi^{'}_{b}$ states, and hope to provide helpful references in theory for the future experiment exploring.

Our theoretical results indicate the newly observed $\Xi^{0}_{b}(6327)$ and $\Xi^{0}_{b}(6333)$ may correspond to
the $\mathbf{\bar{3}}_F$ assignments $\Xi_b|J^P=\frac{3}{2}^+,2\rangle_{\lambda}$ and $\Xi_b|J^P=\frac{5}{2}^+,2\rangle_{\lambda}$, respectively, where the $\Sigma_bK$ channel dominates the decay decay of $\Xi_b|J^P=\frac{3}{2}^+,2\rangle_{\lambda}$, while the partial decay width of the $\Sigma^*_bK$ channel are sizable for $\Xi_b|J^P=\frac{5}{2}^+,2\rangle_{\lambda}$. With this correspondence, $\Xi'_b\pi$ and $\Xi^*_b\pi$ may be another ideal channels for investigating the nature of $\Xi^{0}_{b}(6327)$ and $\Xi^{0}_{b}(6333)$, respectively, in future experiments.

The other two $\rho$-mode $\mathbf{\bar{3}}_F$ states $\Xi_b|J^P=\frac{3}{2}^+,2\rangle_{\rho}$ and $\Xi_b|J^P=\frac{5}{2}^+,2\rangle_{\rho}$ are most likely to be narrow states with a total decay width of $\Gamma\simeq(12-30)$ MeV. $\Xi_b|J^P=\frac{3}{2}^+,2\rangle_{\rho}$ mainly decays into $\Sigma_bK$, while $\Xi_b|J^P=\frac{5}{2}^+,2\rangle_{\rho}$ dominantly decays into $\Sigma^*_bK$. Hence, those two states have a good potential to be observed in their dominant decay process.

For the $1D$ $\Xi'_b$ states belonging $\mathbf{6}_F$, the total decay widths are not broad and vary in the range of $\Gamma\simeq(14-46)$ MeV. Especially for $\Xi'_b|J^P=\frac{3}{2}^+,2\rangle_{\lambda}$ and $\Xi'_b|J^P=\frac{5}{2}^+,2\rangle_{\lambda}$, their total decay widths are about a dozen MeV.

\section*{Acknowledgements }

This work is supported by the National Natural Science Foundation of China under Grants No.12005013, No.11947048, No.12175065, No.U1832173 and No.11775078.

 \end{document}